\documentstyle[prd,aps]{revtex}
\begin{document}
\draft
\title{How much of the  outgoing radiation can be  intercepted
by Schwarzschildean black holes?}

\author{Edward Malec}
\address{Physics Department, University College, Cork,
Ireland and  Institute of Physics, Jagellonian University,
  {3}0-059 Krak\'ow, Reymonta 4, Poland}
\maketitle

\begin{abstract}
  The Schwarzschild spacetime is for electromagnetic waves like
a nonuniform medium with a varying refraction index. A fraction of an
outgoing radiation scatters off the curvature of the  geometry and can be
intercepted by a gravitational center.
The amount of the intercepted energy is bounded above by the  backscattered
energy  of an initially outgoing pulse of electromagnetic
radiation, which in turn  depends on the initial energy,
the Schwarzschild radius and  the pulse location. Its magnitude
depends on the frequency spectrum: it becomes negligible in the short
wave limit but   can be significant in the long wave regime.
\end{abstract}
\pacs{ }
\date{ }

Backscattering   prevents  waves from being transmitted exclusively   along
null cones. That  aspect of   waves propagation has been investigated since
the beginning of XXth century ( see \cite{Hadamard} and, in the context
of general relativity, \cite{MTW}, \cite{collective} \cite{Bardeen1973}).
Electromagnetic waves and  their backscattered  tails
have been studied  from the early seventies \cite{Bardeen1973}.
Much attention  has been put into the explanation of various interference
 phenomena of backscattering tails (\cite{Bardeen1973}, \cite{nil}).

The energy loss in a single burst of radiation
due to this effect has been  assessed only recently
in \cite{Malec00} and (for a scalar field) in \cite{MENOM}.
It should be pointed out that there exist
estimates that refer to a stationary radiation (e. g. Price et al.
in \cite{collective}). It is clear,
that in such a case the backscattered
tails add to the radiation source and the
backscattering effect is underestimated, in some cases  quite significantly.
This paper refines substantially the result of \cite{Malec00}  concerning
the backscattered energy
but the main focus is  on the issue of the dependence of the
effect  on the waves  frequency.

Spherically symmetric geometry outside matter is given by
a  line element,
\begin{equation} ds^2 = - (1-{2m\over R})dt^2 +
{1\over 1-{2m\over R}} dR^2 + R^2 d\Omega^2~,
\label{1}
\end{equation}
where $t$ is a time coordinate, $R$ is a radial
coordinate that coincides with the areal radius
and $d\Omega^2 = d\theta^2 + \sin^2\theta d\phi^2$
is the line element on the unit sphere, $0\le \phi < 2\pi $
and $0\le \theta \le \pi $.

The Maxwell equations read, using a multipole expansion of the
electromagnetic vector potential   \cite{MTW}
\begin{equation}
(-\partial_0^2 + \partial_{r^*}^2)\Psi_l = (1-{2m\over R}){l(l+1)\over R^2}
\Psi_l.
\label{2}
\end{equation}
$\Psi $'s should be  two-index functions, $\Psi_{lM}$,
(where $M$ is the projection of the angular momentum), but since
the evolution equation is $\phi $ independent, the index $M$ is suppressed.
The variable $r^*\equiv R+2m\ln ({R\over 2m}-1)$ is the Regge-Wheeler
tortoise coordinate. The backreaction exerted  by the  electromagnetic
field onto the metric has been neglected in the present analysis. That
assumption is  justified for any gravitational sources other than black holes
while for   black holes  this  approximation holds true  some
 distance away from  its horizon \cite{MENOM}.
In the rest of this paper only  dipole radiation $\Psi_1$
will be considered.   Consequently, all angular momentum related subscripts
will be omitted.

It is  convenient to seek a  solution $\Psi (r^*,t)$ in the form
\begin{equation}
\Psi =\tilde \Psi +\delta ,
\label{3}
\end{equation}
where $\delta $ is an unknown function and
\begin{equation}
\tilde \Psi (r^*,t)= \partial_{r^*} \Bigl( -g(r^*+t)- f(r^*-t)\Bigr) +
{g(r^*+t) +f(r^*-t)\over R}.
\label{4}
\end{equation}
Functions $f$ and $g$ can be uniquely determined from initial data.
One can  check that $\tilde \Psi $  solves Eq. (\ref{2}) in Minkowski
spacetime. The $f$-related   and the $g$-related parts  represent an
outgoing  or  an ingoing radiation, respectively. In what follows
it will be assumed that $g=0$, i. e., that  the wave  is initially
outgoing.

Initially $\delta =\partial_0\delta =0$. Dipole-type initial data
 are given by
\begin{equation}
\tilde \Psi (x(R))=-\partial_{r^*} f(x(R))+{ f(x(R))\over R(r^*)},
\label{5}
\end{equation}
where   $ f$ is  $C^2$-differentiable, has a support  in the annulus
$(a, b\le  \infty )$ and $x(R)
\equiv r^*(R)-r^*(a)$. The initial  energy density  is continuous and
vanishes on the boundary $a$.

The assumption   that initial data are (initially) purely  outgoing is
made  only for the sake of convenience. The propagation of
electromagnetic waves is a linear process as far as the backreaction can be
neglected. Therefore the propagation of the  initially outgoing radiation
(or even of selected modes of the outgoing radiation)
is independent of whether or not  the ingoing radiation (or any other mode)
is present.

The  evolution of $\delta $ is ruled  by
\begin{equation}
(-\partial_0^2 + \partial_{r^*}^2)\delta = (1-{2m\over R})
\Biggl[ { 2\over R^2}
\delta + {6mf\over R^{4}}  \Biggr] .
\label{6}
\end{equation}
The  energy  $E_R(t)$ of the electromagnetic field  $\Psi $
contained in the exterior of a sphere of a radius $R$ reads
\begin{equation}
E_R(t) =2\pi  \int_{R}^{\infty }dr
\Biggl(  {(\partial_0\Psi )^2\over 1-{2m\over r}} + (1-{2m\over r})
(\partial_r\Psi )^2+{2(\Psi )^2\over r^2}\Biggr) .
\label{7}
\end{equation}
 $E_a(0)$ is the energy of the initial pulse.
Let  an outgoing  null cone $C_{a}$ originate from a point $(a,0)$
of the initial hypersurface.
In the Minkowski spacetime the outgoing radiation contained
outside $C_{a}$ does not leak inward and its  energy remains constant.
In a curved spacetime some  energy
 will be lost from the main stream
due to the  diffusion
of the  radiation $h_-$ through   $C_{a}$.  Most of the
backscattered radiation will be intercepted by the gravitational center.

{\bf Theorem.  }  Under the above assumptions,
 the fraction of the diffused energy  $\delta E_a/E_a(0) $
satisfies the inequality
\begin{eqnarray}
&&{\delta E_a\over E_a(0)} \le  C\Bigl( {2m\over a}\Bigr)^2
\Bigl( {1+\sqrt{2m\over a-2m} \over  (1-{2m\over a})^3} \Bigr)^2
(1-{a\over b}),
\label{8}
\end{eqnarray}
where $C$ is a constant depending on $a, m$ and $b$. $C $
decreases with  the increase of $a$ or the decrease of $(b-a)/a$,
and it is bounded - $C<10^2$ (a stricter estimate reads $C<30$).

{\bf  Sketch of the proof. }
Define the intensity of the backscattered   radiation that is directed inward
\begin{equation}
h_-(R,t) ={1\over 1-{2m \over R}}(\partial_0+\partial_{r^*})\delta.
\label{9}
\end{equation}
The rate  of the energy change along $C_{a}$ is given by
\begin{eqnarray}
 (\partial_0+\partial_{r^*})E_{{a}}=
-2\pi (1-{2m\over R})\Biggl[ (1-{2m\over R}) \Biggl( h_--
 { f\over R^{2}}\Biggr)^2
  +{ 2\over R^2}\Bigl( \tilde \Psi+\delta\Bigr)^2 \Biggl] .
\label{10}
\end{eqnarray}
The  energy loss  is equal to a line integral  along     $C_a$
(where $f=\tilde \Psi =0$),
\begin{eqnarray}
  \delta E_a\equiv  E_{a}- E_{\infty }=
2\pi \int_{a}^{\infty  } dr
  \Biggl[ (1-{2m\over r}) h^2_-
  +{ 2 \delta^2\over r^2} \Biggl] .
\label{11}
\end{eqnarray}
The  proof of (\ref{8}) requires the derivation  of estimates
on $h_-$ and $\delta $.  One obtains
\begin{equation}
|{f(R, t=0)\over R}|=|\int_a^R\partial_r{f\over r}|=
|-\int_a^Rdr {\tilde \Psi \over (r-2m}+2m\int_a^Rdr{f\over r^2(r-2m)}|;
\label{11a}
\end{equation}
the Schwarz inequality and the use of (\ref{7}) imply
$|f(R,0)/R|\le \sqrt{aE_a(0)/(4\pi (a-2m))} +2m \int_a^Rdr |f|/(r^2(r-2m)$.

Define $X\equiv \int_a^Rdr {|f|\over r}{1\over r(r-2m)}$;
notice that $X(a)=0$. The  preceding bound of $|f|$
can be written in terms of $X$ as
${dX\over  dR}\le
{\sqrt{E_a(0)/(4\pi)} \over R(R-2m)(1-2m/a)}\sqrt{R-a} +{2mX \over  R(R-2m)}$.
The use of the method of differential inequalities yields
$X \le 2{1-2m/a\over 1-2m/R}\sqrt{E_a(0)/(4\pi}
\Bigl[ {1\over \sqrt{a-2m}}
-{1\over \sqrt{R-2m}}\Bigr] .$

Insertion of that into the  former bound of $f$ gives the  bound,
\begin{equation}
|{f(R,0)\over R}| \le \sqrt{E_a(0)/(4\pi)}\sqrt{R-a}~
{1+\sqrt{2m\over a-2m} \over  1-{2m\over a}}.
\label{11b}
\end{equation}
This bound of $f$  is new but the next steps of the
proof follow quite closely \cite{Malec00}.
i) One obtains an energy estimate on $\delta $, using the energy method and
equation (\ref{6}).
ii) The integration of (\ref{9}) and the use  of i)  yield a  bound
of $h_-$.
iii)  ii) and again (\ref{6})
improve a bound on $\delta $. iv) In bounding the energy loss
due to $\delta $-related terms, one should use both types (i) and iii)
of estimates on $\delta $;  a variational type  argument yields then the
best evaluation of the constant $C$.
The above estimate is sharper than that of
\cite{Malec00}, especially in the regime  $ \kappa \equiv 1-a/b  \approx 1$, when
the bound improves circa 300 times. Details will appear elsewhere.

When the support of the initial radiation is very narrow, i. e.,
$ \kappa  << 1$,  then
${\delta E_a\over E_a(0)} \le C_1\Bigl( {2m\over a}\Bigr)^2\kappa $,
where $C_1$ is a constant.
In the   limit $ \kappa \rightarrow 0$  the ratio
 ${\delta E_a\over E_a(0)}$
becomes 0; the backscattering is negligible when a support of initial pulses
of electromagnetic energy becomes very narrow. And conversely,
the bound becomes bigger with the increase of the width of the radiation
pulse.

The physical meaning of that can be deduced  as follows. Let
$a(t)=a+t+2m \ln ({a(t)\over 2m}-1)$ and
$b(t)=b+t+2m \ln ({b(t)\over 2m}-1)$ be radial components of points lying
on null cones $\Omega_a, \Omega_b$ outgoing from $(a, t=0)$ or $(b, t=0)$,
 respectively. Let $t>>b$; then $b(t)/a(t) \approx 1$.  Then  one can show
that   the energy content between $(a(t), b(t))$
 of the transmitted pulse reads
\begin{equation}
E_{a(t)}(t) \approx
4\pi  \int_{a(t) }^{b(t) }dr
\Bigl( \partial_r^2 f(r)\Bigr)^2
\label{12}
\end{equation}
An auxiliary lemma is needed.

{\bf Lemma.}  Let $f$ be a twice differentiable function,
 $f(b(t)=0$, $R\epsilon (a(t), b(t))$ and $a(t)\approx b(t)$.
Then
\begin{equation}
\int_{a(t)}^{b(t) }dr {f^2\over r^2} <<  \int_{a(t)}^{b(t) }dr
(\partial_rf)^2
\end{equation}
{\bf Proof.} Notice that  $|f/R|=|\int_{b(t)}^Rdr \partial_r(f/r)|$;
that is bounded above (applying the Schwarz inequality and integrating)
 by $\sqrt{1/R}\Bigl( |\int_{b(t)}^Rdr (\partial_rf)^2|\Bigr)^{1/2} +
\sqrt{1/R}\Bigl( |\int_{b(t)}^Rdr  f^2/r^2|\Bigr)^{1/2}$.
Using this one obtains
\begin{equation}
\int_{a(t)}^{b(t) }dr {f^2\over r^2} \le 2\ln {b(t)\over a(t)}
\Biggl( |\int_{b(t)}^Rdr (\partial_rf)^2|+|\int_{b(t)}^Rdr  f^2/r^2|\Biggr) ,
\end{equation}
which, taking into account $\ln b(t)/a(t)\approx 0$,
immediately proves the Lemma.

If $t>>b$ then  terms $2m/r$ can be ignored and (\ref{7}) becomes
 $E_{a(t)}(t)$ of (\ref{12}) plus terms of the form
$\int_{a(t) }^{b(t) }dr(a_0f^2/r^4+ a_1f\partial_rf/r^3+
a_2f(\partial_rf)^2/r^2 +a_3 f\partial_r^2f/r^2 +
a_4 \partial_r^2f\partial_rf/r)$, where
$a_i$ are some constants. But, applying several times
the Lemma, one immediately
shows that all  these terms are much smaller than
$E_{a(t)} $ if $t>>b$ and therefore (\ref{12}) is a valid approximation of the
energy.

>From the Parseval identity follows
\begin{equation}
E_{a(t)}(t)  =
 4\pi  \int_{-\infty ) }^{\infty }dk k^4
|  \hat  f(k)|^2;
\label{13}
\end{equation}
here $\hat f(k)$ is the Fourier transform of $  f(r)$.
The {\it similarity theorem} of the Fourier transform theory
(\cite{Bracewell}) states that compression of the support of a function
corresponds  to expansion of the frequency scale. In explicit terms, if
$a(t)_N\equiv b(t)-(b(t)-a(t))/N$ then the Fourier transform of
$f(r)_N\equiv f(b(t)-N(b(t)-r))$, $\hat f_N(k)$
satisfies    $|\hat f_N(k)|=|\hat f(k/N)|/N$.
 The energy carried by
the rescaled field in modes $\omega \le \Omega_0$ is
$E(\Omega_0)^{(N)}\equiv
4\pi  N^3\int_{-\Omega_0/N ) }^{\Omega_0/N }dk k^4 |  \hat  f(k)|^2$
while the total energy $E_{a(t)}(t)^{(N)}$
is given by  $4\pi  N^3\int_{-\infty  ) }^{\infty }
dk k^4 |  \hat  f(k)|^2=N^3E_{a(t)}(t)$.

The ratio of the two energies
\begin{eqnarray}
 \delta_N(\Omega_0)\equiv   {E(\Omega_0)^{(N)}\over  E_{a(t)}(t)^{(N)}}=
  { \int_{-\Omega_0/N  }^{\Omega_0/N }dk k^4 |  \hat  f(k)|^2
\over E_{a(t)}(t) }
\label{14}
\end{eqnarray}
vanishes in the limit $ N\rightarrow \infty  $.
Thus if a support of initial data is made narrow,  then the wavelengths
scale of the pulse extends in the direction of short lengths, in the sense
that most of the radiation comes in the high frequency band.
That implies, in conjunction with the Theorem, that the high
frequency radiation is essentially unhindered by the effect of
backscattering while  long waves can be backscattered.

It is of interest to determine $\omega_c$ -  a frequency that is critical
in the sense  that  waves  with $\omega \le \omega_c$
may be strongly backscattered while those with $\omega >\omega_c$
can be only weakly backscattered.  As this vague definition suggests,
$\omega_c$ will be determined only up to an order of magnitude.
The bound given in the Theorem shows that there exists a critical width;
if  $b-a$ of an initial pulse is of the order of the distance $a$ from the
gravitational center, then strong backscattering is not excluded, provided in
addition that $a$ is not much greater that the gravitational radius $R_S=2m$.
Thus, in the imprecise sense  of the former definition, the critical width is
$b-a\approx 2m$.  Thus the  sought {\bf  critical frequency } can be
defined as the fundamental  frequency $\omega_c =\pi /R_S$.
One can show, for any  pulse that is smoothly distributed  within an annulus  $(a, b)$,
that most ($\ge 80\%$) of its energy comes with frequencies $\omega \ge \omega_1/2=
\pi /(b-a)$; thus  $\omega_c$ is in fact critical in the sense defined above.

In order to exemplify the above statements, recall estimates of
\cite{Malec00}.  Assume  the same location $a=4R_S$, of two radiative
dipoles and
i)     $ \kappa =1/8$ (i. e., the fundamental wavelength
 $R_S$ is simultaneously critical) for a pulse I;
ii)  $\kappa =1/128$
(i. e., the fundamental wavelength $R_S/8$ is much smaller than the critical
one) for the  pulse II. Then in the case I one obtains
$\delta E_a/E_a(0)<0.37$,
while in the case II (of shorter waves, subcritical case)
one gets  $\delta E_a/E_a(0)<0.001$.
If  the   dipole radiation II is located at $a=4m$ then
 $\delta E_a/E_a(0)\approx 0.77$, which demonstrates how sensitive  the bound
(and presumably the effect itself) is  on the distance.
This dependence of the backscattering  on the wave length has been
observed in the numerical investigation of the propagation of pulses of
scalar massless fields \cite{Regucki}.

The backscattering effect becomes negligible at distances much
bigger than the Schwarzschild radius of a central mass. That rules out most
stars as objects that can  induce observable backscattering effects. For
a star of a solar type and $\lambda \sim R_S$, for instance,   the   ratio
${\delta E_a\over E_a(0)}$  can be at most $10^{-20}$.
In the case of white dwarves and $\lambda \sim R_S$ the above bound gives
${\delta E_a\over E_a(0)}<10^{-8}$. For long-wave radiation the bound
is bigger -   for white dwarves it becomes
 ${\delta E_a\over E_a(0)} \sim 10^{-5}$ - but a sharper
estimate would still lower that  significantly.

Two astrophysical compact objects, neutron stars and black holes, can be  of
interest. They can intercept the  backscattered radiation,
which would possibly lead to the suppression of
the total luminosity produced in accretion
disks that exist in their vicinities. This effect would be  probably weak
since the most luminous regions  of the disks  are located at a distance
of (at least) several Schwarzschild radii.
More interesting can be "echoes" - aftermaths of  flashy eruptions,
produced by a  radiation reflected from the close vicinity of a
horizon of a black hole.
Numerical calculations done in the massless scalar fields propagation
suggest that the amplitude of the reflected long-wave  radiation can
constitute up to 20 \% of the incident one.

The above  results    can be   generalized
into the case of higher order electromagnetic multipoles.
An analysis similar to that of the present paper can be repeated also
in the case of a weak gravitational radiation produced
around Schwarzschildean black holes.

{\bf Acknowledgements.}  This work has been supported in part  by
the KBN grant 2 PO3B 010 16. The author is grateful to Niall O' Murchadha
for     many discussions and valuable comments and to Irene Horne for
the reading of the manuscript. Thanks are due
to members of the Physics Department of UCC for their warm hospitality.

\vfill \eject

\end{document}